\documentclass[11pt]{article}
\usepackage{mathrsfs}
\usepackage{amsmath}
\usepackage{amssymb}
\usepackage{cases}
\usepackage{ulem}
\usepackage{color}
\usepackage{tabularx}

\newcommand{\bF}{ {\mathbb F}}

\newcommand{\EOP} { \hfill $\Box$ }

\newcommand{\pf} { {\rm {\bf Proof.}} }

\newtheorem{theorem}{Theorem}[section]
\newtheorem{prop}[theorem]{Proposition}
\newtheorem{cor}[theorem]{Corollary}
\newtheorem{lem}[theorem]{Lemma}

\newtheorem{rem}[theorem]{Remark}

\newtheorem{example}[theorem]{Example}

\begin{document}
\title{Constructions of involutions over finite fields }
\author{ \scriptsize Dabin Zheng$^{1}$%\footnote{Corresponding author  \quad E-mail: dbzheng@gucas.ac.cn }
\quad  Mu Yuan$^{1}$ \quad Nian Li$^{1}$ \quad  Lei Hu$^{2}$ \quad Xiangyong Zeng$^{1}$\\
\scriptsize 1 \,\, Hubei Key Laboratory of Applied Mathematics, \\
\scriptsize Faculty of Mathematics and Statistics, Hubei University, Wuhan 430062, China \\
\scriptsize 2\,\, Institute of Information Engineering, CAS, Beijing 100093, China  \\
}
\date{}
\maketitle

\begin{abstract}
An involution over finite fields is a permutation polynomial whose inverse is itself.
Owing to this property, involutions over finite fields have been widely used in
applications such as cryptography and coding theory. As far as we know, there are not many
involutions, and there isn't a general way to construct involutions over finite fields.
This paper gives a necessary and sufficient condition for the polynomials of the form $x^rh(x^s)\in \bF_q[x]$
to be involutions over the finite field~$\bF_q$,
where $r\geq 1$ and $s\,|\, (q-1)$. By using this criterion we propose
a general method to construct involutions of the form $x^rh(x^s)$ over $\bF_q$ from given involutions over the
corresponding subgroup of $\bF_q^*$. Then, many classes of explicit involutions of the form $x^rh(x^s)$ over $\bF_q$ are obtained.
\end{abstract}

%%%%%%%%%%%%%%%%%%%%%%%%%%%%%%%%%%%%%%%%%%%%%%%%%%%%%%%%%%%%%%%%%%%%%%%%%%


%%%%%%%%%%%%%%%%%%%%%%%%%%%%%%%%%%%%%%%%%%%%%%%%%%%%%%%%%%%%%%%%%%%%%%%%%%

\section{Introduction}

Let $q$ be a power of a prime. Let $\bF_{q}$ be a finite field with $q$ elements and $\bF_q^*$ denote its multiplicative group.
A polynomial $f(x)\in \bF_{q}[x]$ is called a permutation polynomial if its associated polynomial mapping $f: c\mapsto f(c)$
from $\bF_q$ to itself is a bijection. Moreover, $f$ is called an involution if the compositional inverse of $f$ is itself.
Permutation polynomials over finite fields have been extensively studied due to their wide applications in cryptography, coding
theory and combinatorial design theory. So, finding new classes of permutations with desired properties is of great interest in both theoretical
and applied aspects, and also a challenging task. In many situations, both the permutation polynomials and their compositional
inverses are required. For example, in many block ciphers, a permutation has been used as an S-box for providing confusion
during the encryption process. While decrypting the cipher, the compositional inverse of the S-box comes into the picture.
Thus, it is advantageous for the designer if both the permutation and its compositional inverse have efficient implementations.
This motivates the use of involutions in the S-box of block ciphers. One practical advantage of involutions is that the implementation of the inverse
does not require additional resources, which is particularly useful (as part of a block cipher) in devices with limited resources.

Involutions have been used widely in block cipher designs, such as in AES~\cite{AES}, Khazad, Anubis~\cite{Biryukov2003}
and PRINCE~\cite{Borghoff2012}. For instance, the inverse function over $\bF_{2^8}$ used in the S-box of AES is an involution.
In PRINCE, a linear involution (denoted by $M^\prime$~) was used to ensure $\alpha$-reflection property.
More typically, in Midori~\cite{Banik2015} and iSCREAM~\cite{Grosso2014}, non-linear involutions were used to provide both encryption and decryption
functionalities with minimal tweaks in the circuit. Recently, Canteaut and Rou\'{e} have shown that involutions
were the best candidates against several cryptanalytic attacks by analysing the behaviors of the permutations of an affine equivalent class
with respect to these attacks~\cite{Canteaut2015}. Involutions have been also used to construct Bent functions over finite fields~\cite{CoulterMesnager2017}
and to design codes. For instance, Gallager used an involution (called Gallager's involution transform) to update check nodes
to obtain low-complexity hardware implementation of the sum product algorithm which was used for decoding~\cite{Gallager1963}.

To the best of our knowledge, there are not many known involutions over finite fields.
Recently, Charpin, Mesnager and Sarkar \cite{CharpinMesnagerSarkar2015} firstly discussed Dickson involutions
over finite fields of even characteristic. Then, they \cite{CharpinMesnagerSarkar2016} investigated monomial
involutions and linear involutions, and proposed an approach to constructing involutions from known involutions over
the finite field $\bF_{2^n}$. Rubio et. al. \cite{Rubio2017} further discussed monomial involutions and their fixed points.
Very recently, Fu and Feng~\cite{FuFeng2018} investigated involutory behavior of all known differentially 4-uniform
permutations over finite fields of even characteristic, and most of them were not involutions.
To date, primary constructions of involutions over finite fields seems far from enough.

It is known that every polynomial over $\bF_q$ can be written as $ax^rh(x^s)+b$ for some $r\geq 1$, $s \mid (q - 1)$, $a, b\in \bF_q$ and $h(x) \in \bF_q[x]$\cite{AW2007,Wang2013}.
Due to the importance of the polynomials of the form $x^r h(x^s)$, Wan and Lidl~\cite{WanLidl1991} first gave a criterion for permutation polynomials of this type.
Then, Park and Lee, Wang and Zieve proposed a more concise criterion for permutation behavior of polynomials of the form~$x^r h(x^s)$~\cite{ParkLee2001,Wang2007,Zieve2009},
respectively. By using this criterion, many classes of permutation polynomials of the form $x^rh(x^s)$ have been constructed \cite{AGW2011,DingQuWangYuanYuan2015,GuptaSharma2016,Hou2013-2014,Kyureghyan-Zieve2016,LiHelleseth12016,LiQuChen2017,LiQuWang2017,MaGe2017,TuZengHu2014,Wang2007,ZhaHuFan2017,Zieve2009,Zieve2013}.
But the previous woks were not involved involutory behavior of the polynomials of this type. This paper proposes a criterion for involutions of
polynomials of the form~$x^rh(x^s)$ by the piecewise representation of functions, and a general method to construct involutions of this form
over $\bF_q$ from given involutions over the corresponding subgroup of $\bF_q^*$. Then, many classes of involutions over $\bF_q$ are obtained.

The remainder of this paper is organized as follows. In Section \ref{sec:iff}, we give a necessary and sufficient condition for $x^rh(x^s)$
being an involution over $\bF_q$ and propose a general method to construct involutions of the form $x^r h(x^s)$ over $\bF_q$ from a
given involution over the corresponding subgroup of $\bF_q^*$. Section~\ref{sec:invospecials} presents explicit involutions of the form $x^rh(x^s)$
over $\bF_q$ from the inverse involution over the subgroup of $\bF_q^*$. In Section~\ref{sec:invorelatedpoly} we construct involutions of the form $x^rh(x^s)$
over $\bF_q$ from some special $h(x)$ over $\bF_q$. Section~\ref{sec:subfield} discusses the constructions of involutions over finite fields from
those on their subfields. Finally, we conclude this paper in Section~\ref{sec:concluding}.

\section{A general construction of involutions over finite fields }\label{sec:iff}

Throughout this paper let $q$ be a power of a prime, and $s, d$ be divisors of $q-1$ such that
$sd=q-1$. Let $\bF_{q}$ be a finite field with $q$ elements and $\mu_{d}$ denote the set of $d$th
roots of unity in $\bF_{q}^*$, i.e.,
\begin{eqnarray*}
\mu_{d}=\left\{ x\in \bF_{q}^* \,|\,  x^{d}=1 \right\},
\end{eqnarray*}
which is also the unique cyclic subgroup of $\bF_q^*$ of order $d$. This subgroup can also be represented as $\left\{   x^s \, |\,  x\in \bF_q^*\right\}$.
It is easy to verify that
\[ \bF_q^* = \bigcup_{i=1}^d S_{\alpha_i}, \quad S_{\alpha_i} = \left\{ x\in \bF_q^* \, |\, x^s = \alpha_i \right\},\]
where $\alpha_i \in \mu_d$ for $1\leq i\leq d$. The following well-known lemma gives a necessary and sufficient condition
for $x^rh(x^s)$ being a permutation over $\bF_q$.

\begin{lem}\label{lem:PWZ}\cite{ParkLee2001}\cite{Wang2007}\cite{Zieve2009}
Let $ r, s, d$ be positive integers with $s d =q-1$. Let $h(x)\in \bF_q[x]$. Then $x^rh(x^{s})$ permutes $\bF_q$ if and only if both
\begin{enumerate}
\item[{\rm (1)}]gcd$(r,s)=1$ and
\item[{\rm (2)}]$x^rh(x)^{s}$ permutes $\mu_d$.
\end{enumerate}
\end{lem}

Next we give a criterion for involutions of the form $f(x) =x^r h(x^s)$ for $r\geq 1$, $s| (q-1)$ and $h(x)\in \bF_q[x]$. From above discussions
we can express $f(x)$ in the form of piecewise functions as
  \begin{equation}\label{eq:objf}
  f(x)=x^r h(x^s)=\left\{\begin{aligned}
                       & 0,              & x=0, \\
                       & h(\alpha_1)x^r, & x\in S_{\alpha_1},  \\
                       & \quad \vdots \\
                       & h(\alpha_i)x^r, & x\in S_{\alpha_i}, \\
                       & \quad \vdots \\
                       & h(\alpha_d) x^r, & x\in S_{\alpha_d},
                   \end{aligned}
                   \right.\end{equation}
where $\alpha_i\in \mu_d$, $i=1,2, \ldots, d$. From the piecewise functional representation we obtain
a necessary and sufficient condition for $x^rh(x^s)$ being an involution over $\bF_q$.

\begin{theorem}\label{thm:main}
Let $r, s, d$ be positive integers with $sd=q-1$. Let $g(x)= x^rh(x)^s$ for some polynomial $h(x)\in \bF_q[x]$.
The polynomial $f(x)=x^rh(x^s)$ is an involution over $\bF_{q}$ if and only if
\begin{enumerate}
\item[{\rm (1)}] $r^2\equiv 1 \,\, {\rm mod}\,\,{s}$\,\,  and
\item[{\rm (2)}] $\varphi(z)=z^{\frac{r^2-1}{s}}(h\circ g)(z) h(z)^r=1$ for all $z\in \mu_{d}$.
\end{enumerate}
\end{theorem}
\pf Following notation introduced above, we know that $\bF_q^* = \cup_{i=1}^d S_{\alpha_i}, S_{\alpha_i} = \left\{ x\in \bF_q^* \, |\, x^s = \alpha_i \right\}$,
where $\alpha_i\in \mu_d$. If $x\in S_{\alpha_i}$, i.e., $x^s=\alpha_i$, then
\[ \left( h(\alpha_i) x^r \right)^s = h(\alpha_i)^s x^{sr} = h(\alpha_i)^s \alpha_i^r = g(\alpha_i).\]
So, $ h(\alpha_i) x^r \in S_{g(\alpha_i)}$. From (\ref{eq:objf}) we obtain that
\begin{equation}\label{eq:composition}
f \circ f(x)=\left\{\begin{aligned}
              & 0 ,   & x=0, \\
              & h\left( g(\alpha_1)\right) h(\alpha_1)^r x^{r^2}, &  x\in S_{\alpha_1},\\
              & \quad \vdots \\
              & h\left( g(\alpha_i)\right) h(\alpha_i)^r x^{r^2}, &  x\in S_{\alpha_i}, \\
              & \quad \vdots \\
              & h\left(g(\alpha_{d})\right) h(\alpha_{d})^r x^{r^2}, & x\in S_{\alpha_{d}}.
              \end{aligned}
              \right. \end{equation}

If $r^2\equiv 1 \,\, {\rm mod}\,\,{s}$ and $z^{\frac{r^2-1}{s}}(h\circ g)(z)h(z)^r=1$ for all $z\in \mu_{d}$, then
from (\ref{eq:composition}) we have for $x\in S_{\alpha_i}$,
\[ \begin{split}
 f\circ f(x) &= h\left( g(\alpha_i)\right) h(\alpha_i)^r x^{r^2} \\
             &= h\left( g(\alpha_i)\right) h(\alpha_i)^r x^{ s \frac{r^2-1}{s}} x \\
             &= h\left( g(\alpha_i) \right) h(\alpha_i)^r \alpha_i^{\frac{r^2-1}{s}} x \\
             &=x.
\end{split} \]
This shows that $f$ is an involution over $\bF_q$.

Conversely, if $f$ is an involution over $\bF_q$, then for each $\alpha_i\in \mu_d$ and any $x\in S_{\alpha_i}$, from~(\ref{eq:composition}) we have
\begin{equation}\label{eq:piecewise1}
h(g(\alpha_i))h(\alpha_i)^r x^{r^2} = x.
\end{equation}
Let $\beta$ be another element of $S_{\alpha_i}$, then
\begin{equation}\label{eq:piecewise2}
h(g(\alpha_i))h(\alpha_i)^r \beta^{r^2}=\beta.
\end{equation}
From (\ref{eq:piecewise1}) and (\ref{eq:piecewise2}), we get $(\frac{x}{\beta})^{r^2-1}=1$.
When $x$ runs over $S_{\alpha_i}$, then $\frac{x}{\beta}$ runs over $S_1$. So, $r^2\equiv 1 \,\, {\rm mod}\,\,{s}$.
With this condition, from (\ref{eq:piecewise1}) we have for any $\alpha_i \in \mu_{d}$,
\[1 = h(g(\alpha_i))h(\alpha_i)^r x^{r^2-1} =   h(g(\alpha_i))h(\alpha_i)^r x^{s \frac{r^2-1}{s}} =  h(g(\alpha_i))h(\alpha_i)^r \alpha_i^{\frac{r^2-1}{s}}= \varphi(\alpha_i) .\]
\EOP

\begin{cor}\label{cor:necessity}
Let $f(x)=x^rh(x^s)$ and $g(x) = x^rh(x)^s$ for some positive integer $r$ and $h(x)\in\bF_q[x]$. If $f(x)$ is an involution over $\bF_{q}$, then
$g(x)$ is an involution over the subgroup~$\mu_{d}$.
\end{cor}
\pf It is easy to verify that $x^s\circ f(x)= g(x)\circ x^s$ for any $x\in \bF_q$. If $f(x)$ is an involution over $\bF_q$, then for any $x\in \bF_q$, one has
\[ x^s = x^s\circ \left( f(x)\circ f(x)\right) =  \left( x^s\circ f(x)\right) \circ f(x)= g(x) \circ x^s \circ f(x) = g(x)\circ g(x) \circ x^s. \]
This shows that $g(x)\circ g(x)$ is an identity over $\mu_{d}$, i.e., $g(x)$ is an involution on this subgroup of order~$d$.  \EOP

\begin{rem}
In general, $g(x)$ being an involution over $\mu_d$ can not imply $f(x)$ being an involution over $\bF_q$. But in some special
cases, Theorem~\ref{thm:conversely} in Section~5 presents the similar result on involutory behavior of $x^r h(x^s)$ as that
in Lemma~\ref{lem:PWZ}.
\end{rem}

To obtain involutions of the form $f(x)=x^rh(x^s)$ over $\bF_q$, it is crucial to find a suitable $h(x)$. From Corollary~\ref{cor:necessity},
$g(x)=x^rh(x)^s$ is necessary to be an involution on the subgroup $\mu_{d}$ of $\bF_q^*$ for $f(x)$ being an involution on $\bF_q$.
Next, we give a general method to determine the coefficients of $h(x)$ from a given involution $g(x)$ over $\mu_{d}$.

Let $\alpha$ be a primitive element of $\bF_q$ and $\omega=\alpha^s$ be a generator of the subgroup~$\mu_d$.
Let $\sigma$ be an involution over $\mu_d$, which is represented by
 $$\sigma =\left(\begin{array}{cccccc}
            1            & \omega       & \cdots & \omega^i     & \cdots & \omega^{d-1} \\
            \omega^{\ell_0} & \omega^{\ell_1} & \cdots & \omega^{\ell_i} & \cdots & \omega^{\ell_{d-1}}
          \end{array}\right)
          =\left(\begin{array}{cccccc}
           \omega^{\ell_0} & \omega^{\ell_1} & \cdots & \omega^{\ell_i} & \cdots & \omega^{\ell_{d-1}} \\
           1            & \omega       & \cdots & \omega^i     & \cdots & \omega^{d-1}
          \end{array}\right), $$
where $\ell_0, \ell_1, \ldots, \ell_i, \ldots, \ell_{d-1}$ is a rearrangement of $0, 1, \ldots, d-1$.
Assume that the polynomial $g(x) = x^rh(x)^{s}$ as a mapping on $\mu_d$ is the same as $\sigma$ on it, i.e, for $ 0 \leq i\leq d-1$,
 \begin{equation}\label{eq:gomegai}
 \left\{\begin{aligned}
  g(\omega^i) &=\omega^{ir} h(\omega^i)^{s}  =\omega^{\ell_i}, \\
  \vspace{1cm}
  g(\omega^{\ell_i})& = \omega^{\ell_i r} h(\omega^{\ell_i} )^{s}  =\omega^i.
 \end{aligned}
 \right.
 \end{equation}
Since $\omega = \alpha^s$, from (\ref{eq:gomegai}) there are some $n_i$ and $n_{\ell_i}$ with $0\leq n_i, n_{\ell_i} \leq s-1$ such that
 \begin{equation}\label{eq:homegai}
 \left\{\begin{aligned}
  h(\omega^i) &=\alpha^{dn_i+ \ell_i - ir}, \\
  h(\omega^{\ell_i})& = \alpha^{dn_{\ell_i} + i -\ell_i r} ,
 \end{aligned}
 \right.
 \end{equation}
where $ i =0, 1, \ldots, d-1$ .
By Theorem~\ref{thm:main} and equalities (\ref{eq:gomegai}) and (\ref{eq:homegai}), the polynomial $f(x) = x^r h(x^s)$ is an involution over $\bF_q$
if and only if the following equalities hold:
\[\left\{\begin{aligned}
    r^2 &\equiv  1\,\, {\rm mod}\,\,  s, \\
  \varphi(\omega^i) &= \omega^{i \frac{r^2-1}{s}}h(\omega^{\ell_i}) h(\omega^i)^r = \alpha^{d(n_{\ell_i}+ r n_{i})} = 1, \\
  \varphi(\omega^{\ell_i}) &= \omega^{\ell_i \frac{r^2-1}{s}}h(\omega^{i}) h(\omega^{\ell_i})^r = \alpha^{d(n_i + r n_{\ell_i})}=1 ,\\
 \end{aligned}
 \right. \]
where $i=0, 1, \ldots, d-1$. This is equivalent to
\begin{equation}\label{eq:conditionni}
 \left\{\begin{aligned}
    r^2 \equiv & 1\,\, {\rm mod}\,\, s, \\
 n_{\ell_i}+ r n_i\equiv & 0 \,\, {\rm mod}\,\, s, \\
 \end{aligned}
 \right.
 \end{equation}
since $n_{\ell_i}+ r n_i\equiv 0 \,\, {\rm mod}\,\, s$ is equivalent to
$n_{i}+ r n_{\ell_i} \equiv 0 \,\, {\rm mod}\,\, s$ due to $r^2 \equiv 1\,\, {\rm mod}\,\,s$. It is verified that there exist integers $r, n_i, n_{\ell_i}$
satisfying~(\ref{eq:conditionni}) for $ i =0, 1, \ldots, d-1$. Let $h(x)=\sum_{i=0}^{d-1} h_i x^i\in \bF_q[x]$ be a
reduced polynomial modulo $x^d-1$. Denote by
\[
A = \left(
\begin{array}{cccc}
1 & 1        & \cdots &  1  \\
1 & \omega   & \cdots & \omega^{d-1} \\
1 & \omega^2 & \cdots & \omega^{2(d-1)} \\
  &\cdots    & \cdots &  \\
1 &\omega^{d-1} & \cdots & \omega^{(d-1)(d-1)}
\end{array}
\right),
H = \left(
\begin{array}{c}
h_0 \\
h_1 \\
h_2 \\
\vdots \\
h_{d-1}  \\
\end{array}
\right),
\Lambda = \left(
\begin{array}{c}
\alpha^{dn_0+\ell_0} \\
\vdots \\
\alpha^{dn_i+\ell_i-ir} \\
\vdots \\
\alpha^{dn_{d-1}+\ell_{d-1}-(d-1)r}
\end{array}
\right).
\]
Since $A$ is a Vandermonde matrix, from (\ref{eq:homegai}) the coefficients of $h(x)$ can be uniquely derived from the following linear equations
\begin{equation}\label{eq:hcoefficients}
A H = \Lambda .
\end{equation}
In summary, we have the following theorem.
\begin{theorem}\label{thm:generalconstruction}
Let $\ell_0, \ell_1, \ldots, \ell_i, \ldots, \ell_{d-1}$ be a rearrangement of $0, 1, \ldots, d-1$.
Let $n_i$ be numbers with $0\leq n_i\leq s-1$ for $i=0, 1, \ldots, d-1$. Let~$\alpha$ be a primitive element of $\bF_q$ and $\omega=\alpha^s$ be a
generator of the subgroup $\mu_d$ of $\bF_q^*$. Let $h(x)= h_{d-1}x^{d-1} + h_{d-2} x^{d-2}+\ldots+h_1 x+ h_0\in \bF_q[x]$, whose coefficients are determined
by~(\ref{eq:hcoefficients}). Then $f(x) = x^rh(x^s)$ is an involution over $\bF_q$ if and only if~(\ref{eq:conditionni}) holds.
\end{theorem}

\section{Involutions from the inverse over the subgroup of $\bF_q^*$ }\label{sec:invospecials}

In this section we will give an explicit construction of involutions over $\bF_q$ from the inverse involution over the
subgroup $\mu_d$ of $\bF_q^*$. To this end, we first discuss a special case that $\mu_d$ having at most two elements.
In this case, $s=\frac{q-1}{2}$ and $\frac{1}{2}$ is viewed as the inverse of $2$ modulo $q-1$ if $q$ is even.
Then, $f(x)$ in~(\ref{eq:objf}) is rewritten as
\begin{equation}\label{eq:composition1}
  f(x)=x^r h(x^s)=\left\{\begin{aligned}
                       & 0,          & x=0, \\
                       & a x^r, & x\in S_1,  \\
                       & b x^r,  & x\in S_{-1},
                   \end{aligned}
                   \right.\end{equation}
where $a=h(1)$, $b = h(-1)$ and $S_{\pm 1} = \{c\in \bF_q^* \,|\, c^{\frac{q-1}{2}} =\pm 1\}$.

If $h(1) = h(-1) = a$, then $f(x) = a x^r$ is a monomial over $\bF_q$. Theorem~\ref{thm:main} implies that $f(x) = a x^r$ is an involution over $\bF_q$ if and only if
 \begin{enumerate}
\item[{\rm (1)}] $r^2\equiv 1 \,\, {\rm mod}\,\,{q-1}$,
\item[{\rm (2)}] $a^{r+1}=1$.
\end{enumerate}
This is exactly Proposition~3.1 in~\cite{CharpinMesnagerSarkar2016} if $q$ is even. When $a\neq b$, $f(x)$ in (\ref{eq:composition1}) can be rewritten
as the following form,
\begin{equation}\label{eq:fexpression}
f(x) = \frac{a-b}{2}x^{\frac{q-1}{2}+r}+ \frac{a+b}{2} x^r.
\end{equation}
By Theorem~\ref{thm:main}, we have the following proposition.
\begin{prop}\label{prop:s2q-1}
Let $q$ be a power of an odd prime. Let $\bF_q$ be a finite field and $a, b\in \bF_q$ with $a\neq b$. The polynomial
$f(x)$ in (\ref{eq:fexpression}) is an involution over $\bF_q$ if and only if
\begin{enumerate}
\item[{\rm (1)}] $r^2\equiv 1 \,\, {\rm mod}\,\,{\frac{q-1}{2}}$,
\item[{\rm (2)}] $\frac{a-b}{2} a^{\frac{q-1}{2}+r} + \frac{a+b}{2} a^r =1$ and
$(-1)^r\frac{a-b}{2}b^{\frac{q-1}{2}+r} +\frac{a+b}{2}b^r=(-1)^{\frac{2(r^2-1)}{q-1}}$.
\end{enumerate}
\end{prop}

\begin{example}
In Proposition~\ref{prop:s2q-1}, let $r=1$ and $a=\delta$ and $b=\frac{1}{\delta}$ for some non-square element in $\bF_q$.
It is easy to check that conditions (1) and (2) hold. Then the polynomial $f(x) = \frac{\delta^2-1}{2\delta} x^{\frac{q+1}{2}} + \frac{\delta^2+1}{2\delta}x$  is
an involution over $\bF_q$.
\end{example}

Next, we discuss the case that the subgroup $\mu_d$ of $\bF_q^*$ has at least $3$ elements.
It is clear that the inverse over $\mu_d$ is an involution, which can be represented as
\[ \sigma = \left(
\begin{array}{cccccc}
1 & \omega       & \cdots &  \omega^i     & \cdots  & \omega^{d-1}\\
1 & \omega^{d-1} & \cdots &  \omega^{d-i} &  \cdots & \omega \\
\end{array}
\right).\]
Assume that the polynomial $g(x) = x^rh(x)^{s}$ as a mapping over $\mu_d$ is the same as $\sigma$ on it.
Then the equalities in (\ref{eq:gomegai}) are as follows,
 \begin{equation}\label{eq:inversomegai}
  g(\omega^i) = \omega^{ir} h(\omega^i)^{s}  =\omega^{d-i}, \,\, i = 0, 1, \ldots, d-1 .
 \end{equation}
From (\ref{eq:inversomegai}) we obtain  that
\begin{equation}\label{eq:hexpress}
h(1) = \alpha^{dn_0}\, \,\,{\rm and}\,\,\,  h(\omega^i) = \alpha^{dn_i+d-i-ir}, \,\, i= 1, 2, \ldots, d-1.
\end{equation}
where $n_0, n_1, \ldots, n_{d-1}\in \left\{0, 1, \ldots, s-1\right\}$. Similarly to that in Theorem~\ref{thm:generalconstruction},
the polynomial $f(x) = x^r h(x^s)$ is an involution over $\bF_q$ if and only if the following equalities hold:
\[\left\{\begin{aligned}
 r^2 &\equiv  1\,\, {\rm mod}\,\,  s, \\
 \varphi(1) &= h(1)^{r+1} = \alpha^{dn_0(r+1)} =1,  \\
 \varphi(\omega^i) &= \omega^{i \frac{r^2-1}{s}}h(\omega^{d-i}) h(\omega^i)^r = \alpha^{d(n_i r+ n_{d-i})}=1, \,\, i=1, 2, \ldots, d-1.
 \end{aligned}
 \right. \]
This is equivalent to
\begin{equation}\label{eq:inverseni}
 \left\{\begin{aligned}
    r^2 \equiv & 1\,\, {\rm mod}\,\, s, \\
 n_0(r+1) \equiv & 0 \,\, {\rm mod}\,\, s, \\
 n_1r+n_{d-1}  \equiv & 0 \,\, {\rm mod}\,\, s , \\
  \cdots & \cdots   \\
 n_{\lfloor\frac{d}{2}\rfloor} r+ n_{\lceil \frac{d}{2}\rceil} \equiv & 0 \,\, {\rm mod }\,\, s,
 \end{aligned}
 \right.
 \end{equation}
where $n_0, n_1, \ldots, n_{d-1}\in \left\{0, 1, \ldots, s-1\right\}$.  Let $h(x)=\sum_{i=0}^{d-1} h_i x^i\in \bF_q[x]$ be a reduced polynomial modulo $x^d-1$.
The linear system (\ref{eq:hcoefficients}) becomes
\begin{equation}\label{eq:inversecoefficients}
\left(
\begin{array}{cccc}
1 & 1        & \cdots &  1  \\
1 & \omega   & \cdots & \omega^{d-1} \\
1 & \omega^2 & \cdots & \omega^{2(d-1)} \\
  &\cdots    & \cdots &  \\
1 &\omega^{d-1} & \cdots & \omega^{(d-1)(d-1)}
\end{array}
\right) \left(
\begin{array}{c}
h_0 \\
h_1 \\
h_2 \\
\vdots \\
h_{d-1}  \\
\end{array}
\right) = \left(
\begin{array}{c}
\alpha^{dn_0} \\
\alpha^{dn_1+d-1-r} \\
\alpha^{dn_2+d-2-2r} \\
\vdots \\
\alpha^{dn_{d-1}+1-(d-1)r}
\end{array}
\right).
\end{equation}
In summary, we have the following theorem.
\begin{theorem}\label{thm:fromiverse}
Let $q$ be a power of a prime, $s$ and $d$ be divisors of $q-1$ such that $ds=q-1$.
Let~$\alpha$ be a primitive element of $\bF_q$ and $\omega=\alpha^s$ be a generator of the subgroup~$\mu_d$ of $\bF_q^*$.
Let $h(x)= h_{d-1}x^{d-1} + h_{d-2} x^{d-2}+\ldots+h_1 x+ h_0\in \bF_q[x]$, whose coefficients are determined by the linear system~(\ref{eq:inversecoefficients}).
Then $f(x) = x^rh(x^s)$ is an involution over $\bF_q$ if and only if~(\ref{eq:inverseni}) holds.
\end{theorem}

Assume that $q\equiv 1 \,\, {\rm mod} \,\, 3$ and the subgroup $\mu_d$ of $\bF_q^*$ has exact $3$ elements, i.e, $s=\frac{q-1}{3}$. Then the condition~(\ref{eq:inverseni}) is
reduced to
\begin{equation}\label{eq:conditionn3}
 \left\{\begin{aligned}
    r^2 \equiv & 1\,\, {\rm mod}\,\, \frac{q-1}{3}, \\
 n_0(r+1)\equiv & 0 \,\, {\rm mod}\,\, \frac{q-1}{3}, \\
 n_1r+n_{2}  \equiv & 0 \,\, {\rm mod}\,\,\frac{q-1}{3} .\\
 \end{aligned}
 \right.
\end{equation}
By solving the corresponding linear system (\ref{eq:inversecoefficients}) we get
\begin{equation}\label{eq:hi}
\left\{\begin{aligned}
 h_2 = &\frac{(2+\omega^2) \alpha^{3n_0} - (1+2\omega^2)\alpha^{3n_1+2-r}-(1-\omega^2)\alpha^{3n_2+1-2r}}{3(1-\omega)},  \\
 h_1 = &\frac{(2+\omega)\alpha^{3n_0}- (1+2\omega)\alpha^{3n_1+2-r}-(1-\omega)\alpha^{3n_2+1-2r}}{3(1-\omega^2)},  \\
 h_0 = & \frac{\alpha^{3n_0} +\alpha^{3n_1+2-r}+\alpha^{3n_2+1-2r}}{3}.
 \end{aligned}\right.
\end{equation}
\begin{prop}\label{prop:sq-13}
Let $q$ be a power of a prime with $q\equiv 1 \,\, {\rm mod}\,\, 3$, $\alpha$ be a primitive element of $\bF_q$ and
$\omega=\alpha^\frac{q-1}{3}$ be a primitive cubic root of unity. Let $h(x)= h_2x^2 + h_1 x+ h_0\in \bF_q[x]$, where $h_2, h_1$ and
$h_0$ are given in (\ref{eq:hi}). Then $f(x) = h_2 x^{\frac{2q-2}{3}+r} + h_1 x^{\frac{q-1}{3}+r} + h_0 x^r$ is an involution over $\bF_q$ if and only if (\ref{eq:conditionn3})
holds.
\end{prop}

Let $q=2^{2k}$ for a positive integer $k$ and $\alpha$ be a primitive element of $\bF_q$. Let $\omega=\alpha^\frac{q-1}{3}$ be the cubic root of unity in $\bF_q$.
Set $r=1$. From (\ref{eq:conditionn3}) we get that
$$ n_0=0\,\,\, {\rm and } \,\,\, n_1+n_2 \equiv 0 \,\, {\rm mod}\,\, {\frac{q-1}{3}}.$$
Let $\beta=\alpha^{3n_1+1}$ for $ n_1\in\{ 0, 1, \ldots \frac{q-4}{3}\}$. Substituting these into (\ref{eq:hi}) we get
\[ h_2 = 1+ \omega \beta + \omega^2\beta^{-1}, \,\,\, h_1= 1+\omega^{2} \beta+ \omega\beta^{-1},\,\,\, h_0= 1+ \beta+ \beta^{-1} .\]
From Proposition~\ref{prop:sq-13} we have
\begin{cor}\label{cor:r1}
With notation introduced above, $f(x)=h_2x^\frac{2q+1}{3}+h_1x^\frac{q+2}{3}+h_0x$ is an involution on $\bF_{2^{2k}}$.
\end{cor}

\begin{example}
Let $q=2^8$, $\alpha$ be a primitive element of $\bF_q$ and $\omega=\alpha^\frac{q-1}{3}=\alpha^{85}$ be the cubic root of unity in $\bF_q$.
Set $\beta=\alpha^{3\ell +1}$, where $ \ell \in\{ 0, 1, \ldots \frac{q-4}{3}\}$ and
\[ h_2 = 1+ \omega \beta + \omega^2\beta^{-1}, \,\,\, h_1= 1+\omega^{2} \beta+ \omega\beta^{-1},\,\,\, h_0= 1+ \beta+ \beta^{-1} .\]
Magma verifies that $f_\ell(x)=h_2x^\frac{2q+1}{3}+h_1x^\frac{q+2}{3}+h_0x$ is an involution on $\bF_{2^8}$ for any~$\ell \in\{ 0, 1, \ldots \frac{q-4}{3}\}$.
This experimental result coincides with that of Corollary~\ref{cor:r1}.
\end{example}

Let $q=2^{2k}$ and $\alpha$ be a primitive element of $\bF_q$. Let $\omega=\alpha^\frac{q-1}{3}$ and $r=\frac{q-4}{3}$.
It is easy to verify that integers $n_0, n_1, n_2$ with $0\leq n_i\leq \frac{q-4}{3}$ and  $n_1=n_2$ satisfy~(\ref{eq:conditionn3}).
Substituting these into (\ref{eq:hi}) we get
\[ h_2 = \alpha^{3n_0}, \,\,\, h_1 =  h_0 = \alpha^{3n_0}+ \alpha^{3(n_1+1)} ,\]
where $n_0, n_1\in \{ 0, 1, \ldots \frac{q-4}{3}\}$. From Proposition~\ref{prop:sq-13} we obtain
\begin{cor}\label{cor:rq+13}
With notation introduced above, $f(x)=h_2x^{q-2}+h_0(x^{\frac{2q-5}{3}} + x^{\frac{q-4}{3}})$ is an involution on $\bF_{2^{2k}}$.
\end{cor}

\begin{example}
Let $q=2^8$ and $\alpha$ be a primitive element of $\bF_{q}$. Set
\[ h_2 = \alpha^{3u}, \,\,\, h_0 = \alpha^{3u}+ \alpha^{3(v+1)} \,\, {\rm and} \,\, f_{u,v}(x)=h_2x^{q-2}+h_0(x^{\frac{2q-5}{3}} + x^{\frac{q-4}{3}}),\]
where $u, v\in \{ 0, 1, \ldots \frac{q-4}{3}\}$. Magma verifies that $f_{u,v}(x)$ is an involution over $\bF_q$ for any $u, v\in \{ 0, 1, \ldots \frac{q-4}{3}\}$.
This experimental result coincides with that of Corollary~\ref{cor:rq+13}.
\end{example}

\section{Involutions from monomial mappings over the subgroup of $\bF_q^*$ }\label{sec:invorelatedpoly}

In this section, we will construct more explicit involutions over $\bF_q$ from some special $g(x)\in \bF_q[x]$,
which as a mapping over the subgroup of $\bF_q^*$ is the same as a monomial mapping on it.

\begin{theorem}\label{thm:hsymtric}
Let $q$ be a power of a prime and $r$ be an integer with $r\equiv -1\,\, {\rm mod}\,\, q-1$. Let
$h(x)\in \bF_{q^2}[x]$ satisfy that for any $\beta\in \mu_{q+1}$,
\begin{equation}\label{eq:conditionh}
\beta^{\frac{r^2-1}{q-1}} h(\beta^r) =h(\beta)\in \bF_q .
\end{equation}
If $h(\beta) \neq 0$ for any $\beta\in \mu_{q+1}$, then $f(x) = x^r h(x^{q-1})$ is an involution over $\bF_{q^2}$.
\end{theorem}

\pf It is clear that $r^2 \equiv 1 \,\, {\rm mod }\,\, q-1$. By Theorem~\ref{thm:main}, we only need to show that
$\varphi(\beta) =1$ for any $\beta\in \mu_{q+1}$. Note that $g(\beta) = \beta^r h(\beta)^{q-1} = \beta^r$
and $q-1 \, |\, (r+1)$. From (\ref{eq:conditionh}) we have
\begin{align*}
\varphi(\beta) & = \beta^{\frac{r^2-1}{q-1}} h\left( g(\beta)\right)h(\beta)^r \\
               & =  \beta^{\frac{r^2-1}{q-1}} h( \beta^r )h(\beta)^{r+1} h(\beta)^{-1} \\
               & =  \beta^{\frac{r^2-1}{q-1}} h( \beta^r ) h(\beta)^{-1} \\
               & = 1.
\end{align*}
This completes the proof.   \EOP

\begin{rem}
With notation introduced in Theorem~\ref{thm:hsymtric}, moreover we assume that $\frac{2(r^2-1)}{q-1} \equiv 0 \,\, {\rm mod} \,\, q+1$. The polynomial
$h(x) = \sum_{i\in \Omega}( h_i x^i + h_i^q x^{qi})\in \bF_{q^2}[x]$ satisfies the condition~(\ref{eq:conditionh}), where
\[ \Omega = \left\{  0\leq i\leq q \, |\, \left( \frac{r+1}{q-1} +i\right)(r-1) \equiv 0 \,\, {\rm mod} \,\, q+1 \right\}.\]
\end{rem}

\begin{cor}\label{cor:qb}
Let $q$ be a power of an odd prime, $r=q^2-q-1$ and $h(x)=bx^i+b^qx^{qi}$, where $1\leq i \leq q$. Then
\[f(x)=x^rh(x^s)=x^{q^2-q-1}h(x^{q-1})=bx^{q^2+(i-1)q-1-i}+b^qx^{(1+i)q^2-(1+i)q-1}\]
is an involution on $\bF_{q^2}$ if one of the following condition hold:
\begin{itemize}
  \item $q\equiv 1\,\, {\rm mod}\,\, 4$ and $b$ is an square element in $\bF_{q^2}^*$;
  \item $q\equiv 3 \,\, {\rm mod}\,\, 4$ and $b$ is a non-square element in $\bF_{q^2}^*$.
\end{itemize}
\end{cor}

%Magma has verified Corollary~\ref{cor:qb} for $q=3^k, \,\, k=1, 2, 3, 4$. These experimental results coincide with that of Corollary~\ref{cor:qb}.

Lemma~\ref{lem:PWZ} shows that $x^rh(x^s)$ permutes $\bF_q$ if and only if $\gcd(r, s)=1$ and $x^rh(x)^s$
permutes the subgroup $\mu_{d}$ of $\bF_q^*$, where $d=\frac{q-1}{s}$. So, it is interesting to find the collections of
polynomials which permute $\mu_d$ for certain values of~$d$. Zieve~\cite{Zieve2009,Zieve2013}
proposed a method to construct permutations of the form $x^rh(x)^s$ over $\mu_d$.
Thus, several classes of permutations of the form $x^rh(x^s)$ over $\bF_q$ were obtained.
Following this technique proposed by Zieve in~\cite{Zieve2009,Zieve2013} we obtain many classes of involutions
of the form $x^rh(x^s)$ over $\bF_q$ by careful choices of $h(x)$.

\begin{theorem}\label{thm:specialh}
Let $q$ be a prime power. Let $m$ and $d$ be integers with $d \, |\, \gcd(q-1, m)$. Let $s=\frac{q^m-1}{d}$ and $r$ be an integer
with $r\equiv -1 \,\, {\rm mod} \,\, s$, and $e\equiv \frac{r^2-1}{s} \pmod{d}$ with $e\leq d-1$.
Let $h(x) = \sum_{0 \leq i \leq d-1} h_i x^i \in \bF_q[x]$ be a polynomial with
\begin{equation}\label{eq:hsymtric}
\left\{\begin{aligned}
 h_{e-i}&=h_i, \,\,0 \leq i \leq e;\\
 h_{d+e-i} &=h_i, \,\,e+1 \leq i \leq d-1.
 \end{aligned}
\right.\end{equation}
If $h(x)$ has no root in $\mu_d$, then $f(x) = x^r h(x^s)$ is an involution over $\bF_{q^m}$.
\end{theorem}

\pf It is clear that $r^2 \equiv 1 \,\, {\rm mod }\,\, s$. To show that $f(x)$ is an involution over $\bF_{q^m}$, by Theorem~\ref{thm:main} we only need to
verify  that $\varphi(\beta) =1$ for any $\beta\in \mu_{d}$.

Since $q\equiv 1 \,\, {\rm mod} \,\, d$ we have
\[ \frac{q^d -1}{ q-1}= \sum_{j=0}^{d-1} q^j \equiv  0\,\, {\rm mod} \,\, d. \]
So, $(q-1) \, |\, \frac{q^d-1}{d} $, and $(q-1) \, |\, \frac{q^m-1}{d}$, i.e., $ (q-1) \, |\, s$ since $d\, |\, m$.
In short, $d\, |\, (q-1)$, $(q-1)\, |\, s$ and $s\, |\, (r+1)$. Hence, for $\beta \in \mu_d$ we have $\beta \in \bF_q^*$ and $h(\beta)\in \bF_q$.
Since $h(\beta) \neq 0$, we have $h(\beta)^s = 1$ and $g(\beta) = \beta^r h(\beta)^s = \beta^r = \beta^{-1}$. For $\beta\in \mu_d$,
from (\ref{eq:hsymtric}) we have
 \begin{equation*}\label{eq:halpha}
 \begin{split}
    \beta^{\frac{r^2-1}{s}} h(\beta^{-1}) &= \beta^\frac{r^2-1}{s} \sum_{i=0}^{d-1} h_i \beta^{-i} = \sum_{i=0}^{d-1} h_i \beta^{e-i} \\
                                         &= \sum_{i=0}^{e} h_i \beta^{e-i}  + \sum_{i=e+1}^{d-1} h_i \beta^{e-i} \\
                                         &= h(\beta).
  \end{split}
\end{equation*}
This equality implies that
\begin{align*}
\varphi(\beta) & = \beta^{\frac{r^2-1}{q-1}} h\left( g(\beta)\right)h(\beta)^r \\
               & =  \beta^{\frac{r^2-1}{q-1}} h( \beta^{-1} )h(\beta)^{r+1} h(\beta)^{-1} \\
               & =  \beta^{\frac{r^2-1}{q-1}} h( \beta^{-1} ) h(\beta)^{-1}= 1.
\end{align*}  \EOP

\begin{cor}\label{cor:mdq-1}
Let $q=2^{k}$ for some positive integer $k$ and $m=d=q-1$. Then $s= \frac{q^{m}-1}{d}$ and $\mu_d = \{\alpha^{si}, \,\, |\, \, 0\leq i\leq d-1\}$,
where $\alpha$ is a primitive element of $\bF_{q^{m}}$. If $a, b\in \bF_{q}$ such that $ h(\beta) \neq 0$ for
$h(x)= ax^{q-2} + bx^{q-3} + b$ and any $\beta\in \mu_d$, then $f(x)=x^{s-1}h(x^s)$ is an involution over $\bF_{q^m}$.
\end{cor}
For $q=2^2$, $m=3$ and $r=20$, Magma verifies that $f(x)=x^{20}(\alpha x^{42}+\alpha^2 x^{21}+\alpha^2)=\alpha x^{62}+\alpha^2 x^{41}+\alpha^2 x^{20}$
is an involution over $\bF_{2^6}$. This experimental result coincides with Corollary~\ref{cor:mdq-1}.

\begin{cor}\label{cor:m4d4}
Let $q=3^{2k}$ for some positive integer $k$, and $m=4$, $d=4$. Then $s= \frac{3^{8k}-1}{4}$ and $\mu_4 = \{\alpha^{si}, \,\, |\, \, 0\leq i\leq 3\}$,
where $\alpha$ is a primitive element of $\bF_{3^{8k}}$. If  $a, b, c\in \bF_{3^{2k}}$ such that $ h(\beta) \neq 0$ for $\beta\in \mu_4$,
where $h(x)= ax^3 + bx^2 + ax + c$, then $f(x)=x^{3^{8k}-2}h(x^\frac{3^{8k}-1}{4})$ is an involution over $\bF_{3^{8k}}$.
\end{cor}
For $k=1$, Magma verifies that $f(x)=x^{6559}(\alpha x^{4920}+\alpha^3 x^{3280}+\alpha x+\alpha^5)$
is an involution over $\bF_{3^8}$. This experimental result coincides with Corollary~\ref{cor:m4d4}.

\begin{theorem}\label{thm:byzieve2013}
Let $q$ be a prime power. Let $r, d$ be integers with $r \equiv -1 \,\, {\rm mod}\,\, {q-1}$ and $d \equiv r-1 \,\, {\rm mod}\,\,{q+1}$.
Let $h(x) \in \bF_{q^2}[x]$ be a polynomial of degree $d$ such that $h(0)\neq 0$, and for any $x \in \bF_{q^2}^*$,
\begin{equation}\label{eq:hcondition}
    \left( x^dh(x^{-1})\right)^q=h(x^q).
\end{equation}
Then $f(x)=x^rh(x^{q-1})$ is  an involution on $\bF_{q^2}$ if and only if $h(x)$ has no root in~$\mu_{q+1}$.
\end{theorem}

\pf
If $f(x)$ is  an involution on $\bF_{q^2}$, then $h(x)$ has no root in $\mu_{q+1}$, otherwise $\varphi(z)$ has a root in $\mu_{q+1}$. This is a contradiction by Theorem~\ref{thm:main}.

Conversely, it is verified that $r^2 \equiv 1 \,\, {\rm mod}\,\,{q-1}$ since $r \equiv -1 \,\, {\rm mod}\,\,{q-1}$.
If $h(x)$ has no root in $\mu_{q+1}$, then from (\ref{eq:hcondition}) we have that
\[ h(\beta)^{q-1}=\frac{h(\beta)^q}{h(\beta)}=\frac{h(\beta^{-q})^q}{h((\beta^q)^q)}=\beta^{-d} \]
for any $\beta\in \mu_{q+1}$. So, for $\beta\in \mu_{q+1}$, $g(\beta) = \beta^r h(\beta)^{q-1}=\beta^{r-d}=\beta$ since $d \equiv r-1 \,\, {\rm mod}\,\,{q+1}$.
Set $e=\frac{r+1}{q-1}$. For any $\beta\in \mu_{q+1}$,
\[ \varphi(\beta)  = \beta^\frac{r^2-1}{q-1}h(g(\beta))h(\beta)^{r} = \beta^{e(r-1)}h(\beta)^{r+1}  = \beta^{e(r-1)}h(\beta)^{e(q-1)}  = \beta^{e(r-d-1)}  = 1. \]
Thus, $f$ is an involution on $\bF_{q^2}$ by Theorem~\ref{thm:main}.   \EOP

\begin{rem}
Let $a_i \in \bF_{q^2}$ satisfy $a_0 \neq 0$ and $a_{d-i}=a_i^q$ for $0\leq i \leq d/2$. It is verified that the polynomials $h(x)=\sum_{i=0}^d a_ix^i$
satisfy (\ref{eq:hcondition}).
\end{rem}

\begin{cor}\label{cor:exm}
Let $q$ be a power of a prime and $a\in \bF_{q^2}^*$. The polynomial
$$f(x)= a x^{q^2-3q+1}+a^q x^{q-2}=x^{q-2}(a x^{(q-3)(q-1)}+a^q)$$ is an involution on $\bF_{q^2}$ if and only if one of the following holds:
\begin{itemize}
\item $q \equiv 1 \,\, {\rm mod}\,\,{4}$ and $a^\frac{q^2-1}{2}\neq -1$;
\item $q \equiv 3 \,\, {\rm mod}\,\,{8}$ and $a^\frac{q^2-1}{4}\neq -1$;
\item $q \equiv 7 \,\, {\rm mod}\,\,{8}$ and $a^\frac{q^2-1}{4}\neq 1$.
\end{itemize}
\end{cor}

\pf It is clear that $f(x) = x^{q-2} h(x^{q-1})$, where $h(x)=a x^{q-3} +a^q$, i.e.,
$r=q-2$ and $d=q-3$. It is easy to check that
$h(x)$ satisfies~(\ref{eq:hcondition}). By Theorem~\ref{thm:byzieve2013}, $f$ is an involution on $\bF_{q^2}$ if and only if $h(x)$ has no root in $\mu_{q+1}$.
However, one can verify that $h(x)$ has roots in $\mu_{q+1}$ if and only if $-a^{q-1}$ is in $(\mu_{q+1})^{q-3}$.
So, $h(x)$ has no root in $\mu_{q+1}$ if and only if
\begin{equation}\label{eq:condition}
(-a^{q-1})^\frac{q+1}{{\rm gcd}(q+1,q-3)}\neq 1.
\end{equation}
We discuss this inequality in the following cases.
\begin{enumerate}
\item[{\rm (1)}] If $q$ is even, then $\gcd(q+1, q-3)=1$. We have
$ (-a^{q-1})^{\frac{q+1}{\gcd(q+1,q-3)}} = a^{q^2-1} = 1$.
So, there is no $a\in \bF_{q^2}$ satisfying (\ref{eq:condition}).
\item[{\rm (2)}] If $q \equiv 1 \,\, {\rm mod}\,\,{4}$, then $\gcd(q+1, q-3)=2$. So,
$(-a^{q-1})^{\frac{q+1}{\gcd(q+1,q-3)}} = -a^{\frac{q^2-1}{2}}$.
When $a^\frac{q^2-1}{2}\neq -1$, then (\ref{eq:condition}) holds.
\item[{\rm (3)}] If $q \equiv 3 \,\, {\rm mod}\,\,{8}$, then $\gcd(q+1, q-3)=4$. Thus,
$ (-a^{q-1})^{\frac{q+1}{\gcd(q+1,q-3)}}  = -a^{\frac{q^2-1}{4}}$.
When $a^\frac{q^2-1}{4}\neq -1$, then (\ref{eq:condition}) holds.
\item[{\rm (4)}] If $q \equiv 7 \,\, {\rm mod}\,\,{8}$, then $\gcd(q+1, q-3)=4$. Hence,
$(-a^{q-1})^{\frac{q+1}{\gcd(q+1,q-3)}}  = a^{\frac{q^2-1}{4}}$.
When $a^\frac{q^2-1}{4}\neq 1$, then (\ref{eq:condition}) holds.
\end{enumerate}
\EOP

\begin{theorem}\label{thm:symetric}
Let $q$ be a power of a prime~$p$ and $d$ be an integer with $q\equiv -1 \,\, {\rm mod}\,\,d$. Let $m$ and $s$ be integers
satisfying $q^{2m}-1=s d $. The polynomial $f(x)=x(1+x^s+\ldots +x^{(k-1)s})$ is an involution on $\bF_{q^{2m}}$
if the following conditions hold:
\begin{equation}\label{eq:degreecond}
 \left\{\begin{aligned}
 (k-1)\gcd\left(k+1, \frac{m(q^2-1)}{2d} -1\right) & \equiv 0 \,\, {\rm mod}\,\, d, \\
 k^2 & \equiv 1 \,\, {\rm mod}\,\, p.
 \end{aligned}
 \right.
 \end{equation}
\end{theorem}
\pf From our assumption we have $s =\frac{q^2-1}{d} \left( \sum_{i=0}^{m-1} q^{2i}\right)$. So, $(q-1) \, |\, s$.
Set $h(x)=1+x+x^2+\ldots+x^{k-1}$, and so $f(x)= xh(x^s)$. Since $d \mid q+1$, for any $\beta\in \mu_d $, $\beta^q = \beta^{-1}$ and
\begin{equation}\label{eq:symtric}
\left( \frac{h(\beta)}{\beta^{(k-1)/2}}\right)^{q} =\left( \beta^{-\frac{k-1}{2}}+\beta^{-\frac{k-3}{2}}+\ldots + \beta^{\frac{k-1}{2}}\right)^q =\beta^{\frac{k-1}{2}}+\ldots +\beta^{-\frac{k-3}{2}}+ \beta^{-\frac{k-1}{2}}.
\end{equation}
If $k$ is even, then $d$ is odd from the first congruent equality in~(\ref{eq:degreecond}). In this case, $\frac{1}{2}$ is viewed as the inverse of $2$ modulo $d$.
The equality (\ref{eq:symtric}) shows that $\frac{h(\beta)}{\beta^{(k-1)/2}}\in \bF_q$. So, we have
 \[ h(\beta)^{q-1} = \beta^{\frac{(k-1)(q-1)}{2}} \,\,\, {\rm and }\,\,\, h(\beta)^{q^2} = h(\beta) .\]
These further imply that
\[g(\beta) = \beta h(\beta)^s=\beta h(\beta)^{\frac{q^2-1}{d}\left( \sum_{i=1}^{m-1} q^{2i}\right)}=\beta h(\beta)^{\frac{(q^2-1)m}{d}} = \beta^{\frac{m(k-1)(q^2-1)}{2d}+1} .\]

To show that $f(x)$ is an involution on $\bF_{q^{2m}}$, by Theorem~\ref{thm:main} it is sufficient to verify that for any $\beta\in \mu_d$,
$\varphi(\beta)=h(g(\beta))h(\beta)=1$. By the first congruence of (\ref{eq:degreecond}) we know that $d\, |\, (k^2-1)$ and $d \, |\, (k-1)\left( \frac{m(q^2-1)}{2d}-1\right)$.
For $1\neq \beta\in \mu_d$,
\begin{align*}
\varphi(\beta) & = h(g(\beta))h(\beta) = h\left(\beta^{\frac{m(k-1)(q^2-1)}{2d}+1}\right)h(\beta) \\
               & = \frac{\beta^{\frac{m(k-1)(q^2-1)+2d}{2d}k}-1}{\beta^{\frac{m(k-1)(q^2-1)+2d}{2d}}-1} \cdot \frac{\beta^{k}-1}{\beta-1} \\
               & = \frac{\beta^{k^2}-1}{\beta^k-1} \cdot \frac{\beta^{k}-1}{\beta-1}\\
               & = 1.
\end{align*}
For $\beta=1$, from (\ref{eq:symtric}) we know that $k=h(1)\in \bF_q^*$. So, $g(1) = h(1)^s = k^{(q-1) \frac{s}{q-1}} \equiv 1 \,\, {\rm mod}\,\, p$.
By the second congruence of (\ref{eq:degreecond}), we have $\varphi(1) =h(g(1)) h(1) \equiv k^2 \equiv 1 \,\, {\rm mod}\,\, p$.   \EOP

\begin{example}
Let $q=3^2, d = 5, m=4$ and $k=4$. Then $s=(q^m-1)/d = 1312$. It is easy to verify that these parameters satisfy
(\ref{eq:degreecond}). By Theorem~\ref{thm:symetric} we know that $f(x)=x+x^{1313}+x^{2625}+x^{3937}$ is an involution on $\bF_{3^8}$.
This is also verified by Magma.
\end{example}

\section{Involutions from known ones over the subfields}\label{sec:subfield}

Let $f(x)=x^rh(x^s)$ and $g(x)=x^rh(x)^s$ be polynomials over $\bF_q$. It is necessary that $g(x)$ must be an involution over the subgroup $\mu_d$
of $\bF_q^*$ for $f(x)$ being an involution over $\bF_q$. Generally, the converse is not true.
But it is still true under some additional conditions. In this section, we construct new involutions over finite fields from
known ones on their subfields.

\begin{theorem}\label{thm:conversely}
Let $q$ be a prime power and $r, s$ be integers with $r^2 \equiv 1 \,\, {\rm mod}\,\,{s}$ and $\gcd(s, d )=1$ for $d=\frac{q-1}{s}$.
Let $h(x)\in \bF_q[x]$ satisfy $h(\mu_d)\subset \mu_d$. Then $f(x)=x^rh(x^s)$ is an involution on $\bF_{q}$ if and only if
$g(x)=x^rh(x)^s$ is an involution on $\mu_d$.
\end{theorem}

\pf
By Corollary~\ref{cor:necessity}, $g(x)=x^rh(x)^s$ is an involution on $\mu_d$ if $f(x)=x^rh(x^s)$ is an involution on $\bF_{q}$.
Conversely, by the Theorem~\ref{thm:main}, it is sufficient to show that $\varphi(\beta)=1$ for any $\beta\in \mu_d$.
Since $h(\mu_d)\subset \mu_d$, we have that $\varphi(\beta)\in \mu_d$ for any $\beta\in \mu_d$.
If $g(x)$ is an involution on $\mu_d$, then for $\beta\in \mu_d$,
\[\begin{split}
\left( \varphi(\beta) \right)^s &= \beta^{r^2-1} h\left(g(\beta)\right)^s h(\beta)^{sr} \\
                                 &= \beta^{-1}h\left(g(\beta)\right)^s\left( \beta^r h(\beta)^s\right)^r \\
                                 &=  \beta^{-1 }h\left(g(\beta)\right)^s g(\beta)^r \\
                                 &=\beta^{-1}g\circ g(\beta) \\
                                 &=1 .
\end{split}\]
Thus $\varphi(\beta)=1$ for any $\beta\in \mu_d$ since $\gcd(s,d)=1$.
\EOP

If we choose a special $s$ in Theorem~\ref{thm:conversely} such that $\mu_{d}$ is exactly a multiplicative group of a finite field,
then the following corollary is obtained.

\begin{cor}\label{cor:appofconversely}
Let $q$ be a prime power and $m, r$ be positive integers with ${\rm gcd} (q-1,m)=1$ and $r^2\equiv 1 \,\, {\rm mod}\,\,{\frac{q^m-1}{q-1}}$. Let $h(x)\in \bF_q[x]$.
Then $f(x)=x^rh(x^\frac{q^m-1}{q-1})$ is an involution on $\bF_{q^m}$ if and only if $g(x)=x^rh(x)^m$ is an involution on $\bF_q$.
\end{cor}
\pf Let $s=\frac{q^m-1}{q-1}$. Since $s= m+\sum_{i=1}^{m-1}(q^i-1) \equiv m \,\, {\rm mod}\,\,{q-1}$ and $h(x)\in \bF_q[x]$, we have that $\gcd(q-1,s)=\gcd(q-1,m)=1$
and $g(x)=x^rh(x)^\frac{q^m-1}{q-1}=x^rh(x)^m$ for any $x\in \bF_q$. By Theorem~\ref{thm:conversely}, $f(x)=x^rh(x^\frac{q^m-1}{q-1})$ is an involution on
$\bF_{q^m}$ if and only if $g(x)=x^rh(x)^m$ is an involution on $\bF_q$.  \EOP

Corollary~\ref{cor:appofconversely} allows us to construct new involutions over finite fields from the known ones on their subfields.

\begin{example}
Let $q=2^{2k}$ for a positive integer~$k$. From Corollary~\ref{cor:r1} we know that
$h_2x^{\frac{2q+1}{3}} + h_1 x^{\frac{q+2}{3}}+ h_0x$ is an involution over $\bF_q$, where
$h_2, h_1, h_0$ are introduced in Corollary~\ref{cor:r1}. Assume that $h(x)\in \bF_q[x]$ satisfies
\[xh(x)^2 = h_2x^{\frac{2q+1}{3}} + h_1 x^{\frac{q+2}{3}}+ h_0x .\]
By Corollary~\ref{cor:appofconversely} we have that $f(x) = xh(x^{q+1}) = h_2^\frac{1}{2}x^\frac{q^2+2}{3}+h_1^\frac{1}{2}x^\frac{q^2+5}{6}+h_0^\frac{1}{2}x$
is an involution over $\bF_{q^2}$.
\end{example}

\begin{example}
Let $q=3^k$ for a positive integer $k$ and $a\in \bF_{q^2}^*$.
Let $h(x)=a^\frac{1}{3}x^\frac{q^2-3q}{3}+a^\frac{q}{3}x^\frac{q-3}{3}\in \bF_{q^2}[x]$ and $f(x)=xh(x^{q^4+q^2+1})$.
It is clear that $\gcd(3, q^2-1)=1$. By Corollary~\ref{cor:appofconversely}, $f(x)$ is an involution on $\bF_{q^6}$ if and only if
 $g(x)=xh(x)^3=a x^{q^2-3q+1}+a^q x^{q-2}$ is an involution on $\bF_{q^2}$. From Corollary~\ref{cor:exm} we have that
the polynomial
$$ f(x)= a^\frac{1}{3}x^\frac{q^6-3q^5+q^4-3q^3+q^2-3q+3}{3}+a^\frac{q}{3}x^\frac{q^5-3q^4+q^3-3q^2+q}{3} $$
is an involution on $\bF_{q^6}$ if and only if one of the following hold:
\begin{itemize}
\item $k$ is even and $a^\frac{q^2-1}{2}\neq -1$;
\item $k$ is odd and $a^\frac{q^2-1}{4}\neq -1$.
\end{itemize}
\end{example}

\section{Concluding remark }\label{sec:concluding}

In this paper, we have systematically studied involutory behavior of the polynomials over $\bF_q$
of the form $x^rh(x^s)$ for some divisor $s$ of $q-1$. A necessary and sufficient condition for $x^rh(x^s)$
being an involution has been obtained. From this criterion we have proposed a general method to construct involutions of the form $x^rh(x^s)$
over $\bF_q$ from given involutions over the corresponding cyclic subgroup $\mu_d$ of $\bF_q^*$.
Then, many classes of explicit involutions over $\bF_q$ have been obtained from involutions over the small
subgroup $\mu_d$ or monomial mappings over some general $\mu_d$. The proposed method can be used to find more desired involutions and
the obtained involutions may have direct applications.

%%%%%%%%%%%%%%%%%%%%%%%%%%%%%%%%%%%%%%%%%%%%%%%%%%%%%%%%%%%%%%%%%%%%%%%%%%%%%%%%%%%

\end{document}